# Information Systems and Software Engineering: The Case for Convergence

Brian Fitzgerald,
Lero – the Irish Software Engineering Research Centre,
University of Limerick, IRELAND

**Abstract**
The Information Systems (IS) and Software Engineering (SE) fields share a remarkable number of similarities in their historical evolution to date. These similarities are briefly outlined below. An analysis of 10 years (2001-2010) of publications in the primary journals in both fields also reveals a good deal of overlap in research topics. Given the challenges faced by both as young disciplines, there is potentially much to gain from a closer interaction between both fields than has traditionally been the case. This article seeks to encourage such interaction, and illustrates how this might usefully occur in the area of design. It concludes by proposing a number of practical initiatives that could stimulate and facilitate interaction between the IS and SE fields.

**Introduction**
There are many similarities between the Information Systems and Software Engineering fields. Both have had a similar trajectory of evolution to date. For example, both fields are about 50 years old, and tend more towards the applied rather than basic science. Both faced significant challenges and questions early on in their existence. Both have two primary journals which are very well respected, and each has a well-established annual international conference which has run for many years. Both have a world-wide mailing list/portal which is used to disseminate information to the global community. Both continue to face challenges within the larger academic disciplines and departments in which they typically reside. In the main, Software Engineering has arisen from within in Computer Science or Engineering departments, whereas Information Systems has come from Business/Management and sometimes Computer Science departments. The disciplines are linked in the public mind as student applications to undergraduate courses in both areas can fluctuate in a similar pattern, although employers typically report a shortage of skilled graduates in the ICT sector generally. Given the similarity of evolution and the challenges being faced by both disciplines, this paper argues that each could benefit from a closer interaction with the other than has traditionally been the case.

Below we sketch the evolution of each field. In doing so, the similarities immediately become obvious. To provide some empirical grounding we analyze the publication patterns over the past 10 years in the two primary journals in each field and consider the extent of potential overlap between the disciplines. We then take a particular example, systems design, and illustrate how a convergent approach which includes insights from both fields could result in a very useful outcome. Finally, we suggest ways in which a fruitful interaction between both fields might occur in a practical fashion.

**Brief Outline of Evolution of IS Field**

Early work which helped define the IS field includes Borge Langefors (1966) infological equation:

> $I = i(D,S,t)$
> where *I is the information (or knowledge) produced from the data D, and the recipient's prior experience (world view) S, by the interpretation process i, during the time period t.*

A key insight here is that given the inherent variety in S, different individuals will differ in the information interpreted even from simple data. This basic principle has been borne out in much subsequent IS research.

Other fundamental work outlining the theoretical and conceptual foundations of IS was published by Gordon Davis (1974) who clearly identified the man/machine interaction at the heart of IS. Over many years of IS research it is now well established that the IS field focuses on the interaction between the technological elements (hardware, software etc) and the social elements (people, processes, organization culture, power and politics etc). Buckingham (1987) provides a key insight in proposing that an information system is a human activity (social) system which may or may <u>not</u> involve the use of computer systems.

Davis (2000) also usefully identified three generic approaches to establishing conceptual foundations for a discipline: an *intersection* approach which accepts concepts from any field that appear to add insight; a *core* approach that defines the ideas that characterize a discipline and make it distinct; and an *evolutionary* approach that combines the core approach with concepts from other fields that over time are found to be useful. In the early years of the field, the intersection approach was the predominant one, and over time a core approach has been attempted. Davis suggests that the evolutionary approach will become more prevalent in the future.

Ahituv and Neumann (1986) define the IS field as follows:

> " the systematic study of information systems. An information system is a set of components (people, hardware, software, data, and procedures) that operate together to produce information that supports the operation and management functions of an organization'

In keeping with this, more recent research (Sidorova et al. 2008) analyzed a body of IS research papers published over a 22 year period and concluded that the IS discipline focuses on how IT systems are developed and how individuals, groups, organizations, and markets interact with IT.

**The Sabre Case**

An early and very well-known exemplar of an information system was SABRE which was developed and installed in 1960 as a joint venture between IBM and American

Airlines. SABRE was a data processing system that could create and manage airline seat reservations, and instantly make that data available electronically to any agent at any location. The SABRE acronym stands for Semi-Automated Business Research Environment[1], and was an evolution of a 1950s military project, SAGE (Semi-Automated Ground Environment). Interestingly, the military project was considered a much more complex endeavor – to the extent that SABRE was known by its developers as "kiddie's SAGE". The concepts underpinning SAGE (and SABRE) were very much ahead of their time including early variants of the systems development life-cycle, the separation of analysis, design and programming, object orientation, and decision support systems. In 1960, SABRE was estimated to be processing 84,000 calls per day. By 1964, it was the largest private transaction processing system and estimated as saving 30% in staff costs alone for American Airlines. By 1974, SABRE was spun off by American Airlines and was deployed widely in travel agents where it was handling 1 million fares per day.

Despite these impressive credentials however, the success of SABRE was not widely replicated. Critiques of the fledging IS field appeared early. In 1967, Russell Ackoff published a seminal paper concerning what he termed Management *Mis*-Information Systems. He outlined five key assumptions of MIS designers which he claimed were erroneous. Table 1 summarizes these assumptions and Ackoff's counter-view of key problems (Ackoff 1967).

**Table 1 Ackoff's *Mis*-Information Systems Tenets (1967)**

| Assumption | Ackoff 's Counter-View |
|---|---|
| 1. Management needs more information | While management lack vital information, a greater malaise is the overabundance of irrelevant information |
| 2. Managers need the information they want | Management tend not to have an adequate model for decision making and thus play it safe by requesting as much information as possible from MIS designers |
| 3. Giving managers the information they need improves their decision making | Because of the complexity of the decision process, managers typically cannot use the information provided well |
| 4. More communication means better performance | Organizational sub-units often have conflicting performance measures and thus more communication may hurt organizational performance, not help it. |
| 5. Managers need only to understand how to use an information system | Managers must understand the information systems they rely on and hence control these systems rather than be controlled by them. |

---

[1] Also referred to as SABER (Semi-Automated Business Environment Research) in some publications

It is a tribute to Ackoff's insight that Most of the issues identified by him in 1967 remain quite problematic to this day in the IS field.

The first annual international conference on information systems (ICIS) was held in 1980 in Philadelphia. Now an established event for the field with a relatively fixed date every December, it has been hosted on three continents (not yet having been in South America or Africa), and typically attracts between 1,000 and 1,500 delegates annually.

The ISWorld web portal, which facilitates communication among members of the IS discipline throughout the world, was established in 1996.

There are two principal journals serving the IS field. *MIS Quarterly (MISQ)*, published by University of Minnesota, was established in 1977. It has become a very influential journal in the ICT domain overall. Its ISI impact factor is 4.485 and its five-year impact factor is 9.208. A second journal, *Information Systems Research (ISR)*, published by INFORMS (Institute for Operations Research and the Management Sciences) was established in 1990. Its ISI impact factor is 3.358 and its five-year impact factor is 5.458. This is also a very reputable and commendable achievement. In ISI's IS & CS combined categories these are two of the journals with the highest impact factors.

In terms of research impact, IS can justifiably claim to have initiated or significantly developed a number of important and influential research topics, including *decision support systems*, *soft systems methodology, team coordination and management, technology acceptance model* and *design science*.

**Brief Outline of Evolution of SE Field**
Early pioneers in the SE field include David Parnas whose work on modularity and information hiding (Parnas 1972) paved the way for object-orientation later, discussed again below. (Interestingly, the Simula programming language which was one of the first embodiments of object-oriented concepts was greatly influenced by Langefors' work which, as already mentioned, was foundational in IS). Edsger Dijkstra's (1968) Go To Statement Considered Harmful paper advocated the use structured programming constructs. The structured movement subsequently evolved from structured programming to focus in turn on structured design and structured analysis (Ward 1991).

Fred Brooks published *The Mythical Man Month* (1975) to document his experiences in developing the IBM OS360 operating system – the latter was reckoned to be the most complex thing that mankind had ever created up to that point. Given such complexity, it is not surprising that a more disciplined engineering-like approach to software development would be an attractive proposition. In the interim, the complexity of software systems has increased not diminished.

While there does not appear to have been as much emphasis on developing conceptual foundations for the field as a whole in SE, one fairly high profile initiative has been the development of the SWEBOK (software engineering body of knowledge). Leaning towards Davis's (2000) *core* approach to establish conceptual foundations, SWEBOK defines software engineering as:

> "the application of a systematic, disciplined, quantifiable approach to the development, operation, and maintenance of software, and the study of these approaches; that is, the application of engineering to software."

Another definitional initiative to coalesce software engineering knowledge in software engineering is the "Computing Curriculum Software Engineering (CCSE)", officially named Software Engineering 2004 (SE2004). While SWEBOK defines the software engineering knowledge that practitioners should have after four years of practice, SE2004 defines the knowledge that an undergraduate software engineering student should possess upon graduation (including knowledge of mathematics, general engineering principles, and other related areas).

One of the earliest recorded uses of the term, software engineering, was at a Nato-sponsored conference in Garmisch in 1968 (Naur and Randell 1968). Just as in IS, critiques in the fledging SE field also appeared early. The term 'software crisis' was coined at the same Nato conference to refer to the problems that were already being perceived in that software was taking too long to develop, cost too much and was not of adequate quality when delivered.

The international conference on software engineering (ICSE) conference series began in 1975 in Washington DC. It is now well established in a May/June calendar slot, and has been hosted on four continents[2]. It is jointly sponsored by IEEE-CS and ACM Sigsoft and attracts about 900 to 1,000 delegates each year.

The SEWorld web portal for communication to members of the SE discipline throughout the world was established in 1999.

There are two principal journals serving the SE field. *IEEE Transactions on Software Engineering (TSE)* was established in 1975 and has an ISI impact factor of 3.75 and a five-year impact factor of 4.86. In a fairly similar evolution to that of IS, a second journal *ACM Transactions on Software Engineering and Methodology (TOSEM)* was established in 1992. *TOSEM* has an ISI impact factor is 2.03 and its five-year impact factor is 3.77. In the overall ISI ranking of ICT journals these two are the highest SE-related journals.

High impact research topics that have emerging primarily from software engineering include concepts such as *information hiding, cohesion* and *coupling* (these provide a basis for *object-orientation*), *software architecture*, *software as a service (SaaS)*, *Software Product Lines*.

Table 2 summarizes the similar evolution of the IS and SE fields as described above.

---

[2] A conference planned for Argentina in South America in 2002 was relocated to Florida due to political unrest in Argentina at the time

## Table 2 Similarities in Evolution of IS and SE Fields

|  | IS | SE |
|---|---|---|
| **Origins of Field** | 1950/60s | 1950/60s |
| **Early Challenges** | Ackoff (1967) | Garmisch (1968) |
| **Primary Journals** | *MISQ* (1977) Impact 4.485<br>*ISR* (1990) Impact 3.358 | *TSE* (1975) Impact 3.75<br>*TOSEM* (1992) Impact 2.03 |
| **Conferences** | ICIS (1980) | ICSE (1975) |
| **Web Portal** | ISWorld (1996) | SEWorld (1999) |

**Analysis of Ten Years of Publication in IS and SE Journals**

The principal topics covered in *MISQ* and *ISR* in the ten-year period, 2001-2010, are listed in Table 3. As can be seen from the table, this involved 352 articles in *MISQ* and 262 articles in *ISR*. A weighted keyword analysis was then used to identify the top 10 topics in terms of extent of coverage in both journals in the ten-year period.

## Table 3 Topic Coverage in *MISQ* and *ISR* articles 2001-2010

| MISQ (352) | ISR (262) |
|---|---|
| Technology Acceptance (8%) | E-Commerce (6%) |
| Research Method (5%) | Privacy/Security/Trust (3%) |
| IS Development (5%) | Technology Acceptance (3%) |
| Knowledge Management (4%) | IS Development (3%) |
| Groups/Teams (3%) | Decision Support Systems (3%) |
| E-Commerce (3%) | Team * (2%) |
| Web/Internet (2%) | Outsourcing (2%) |
| Outsourcing (2%) | Web/Internet (2%) |
| ERP/Enterprise Systems (2%) | Agile (2%) |
| Security (2%) | Computer-Mediated Communication (2%) |

As can be seen from Table 3, there is quite a lot of overlap in topic coverage across both journals, albeit slightly different rankings across each. The *Team *** convention was chosen for the *ISR* analysis to cluster a range of team-related research issues such as *team formation, team performance, team conflict, virtual teams*.

Several other topics were close to the threshold for inclusion in the top 10. These include *decision support systems* and *trust* in *MISQ* and, at a slightly lower level *open source*

*software* and *sustainable energy/environment*. In the case of *ISR*, topics close to the top 10 include *business value of IT* and *open source software*.

Of interest in the *MISQ* and *ISR* analysis is the extent of fragmentation. This is well captured by the fact that the top 10 topics account for only 36% of the publications in *MISQ* and 28% in *ISR*.

Table 4 below presents an analysis of the topic coverage in the two primary SE journals – TSE (685 articles) and TOSEM (147 articles) for the period 2001-2010. Again, the same weighted keyword analysis was used to derive the top 10 topics in terms of popularity in the period.

**Table 4 Topic Coverage in *TSE* and *TOSEM* articles 2001-2010**

| TSE (685) | TOSEM (147) |
|---|---|
| Testing/Debugging/Defects (8%) | Design (20%) |
| Model * (4%) | Verification (13%) |
| Analysis/Design/Development (3%) | Languages (10%) |
| Object-Oriented * (3%) | Algorithms (10%) |
| Requirements * (2%) | Reliability (4%) |
| Formal * (2%) | Experimentation (4%) |
| Distributed Development (2%) | Security (3%) |
| Cost Estimation (2%) | Management (3%) |
| Empirical Software Engineering (2%) | Measurement (3%) |
| Real-Time * (2%) | Theory (3%) |

Of interest is the apparent fragmentation in *TSE* where the top 10 topics account for just 30% of the articles. To achieve even this result, some clustering was necessary. We group similar topics using a '*topic \**' conventions. Thus *model \** in the table above refers to a clustering of model-related topics such as *modelling*, *model formalisms, model checking, modelling methodologies, modelling tools, modelling techniques, modelling frameworks*. This approach was also adopted for several other topics – *object-oriented \*, requirements \*, formal \*, and real-time \**. Topics that were close to the threshold for inclusion include *software architecture, security*, and *web \**.

In the case of *TOSEM*, keyword usage patterns were much less fragmented. We attribute this primarily to the fact that *TOSEM* submissions use the ACM Computing Classification System (CCS) keyword scheme (http://www.acm.org/class/1998) on article submission. Thus the top 10 topics in *TOSEM* account for 73% of the papers. Topics that approached the inclusion threshold for *TOSEM* included *documentation* and *human factors*.

**Topic Overlap in IS and SE Journals**
There is evidence of overlap across the two disciplines, most strikingly in the area of

*systems development*. In both IS journals this features as a major research topic in its own right with strongly complementary research topics such as *agile*, *groups* and *team \**. In the case of the SE journals, *design/development* features with complementary research topics such as *requirements \*, model \*, object-oriented \** and *testing*. *Security* is also a topic that was common across the four journals.

Fig 1 suggests how IS and SE research tend broadly speaking to focus attention on different parts of the systems development life-cycle. While this is a rather crude generalization for the sake of simplicity, these research foci are very complementary. IS focuses relatively heavily on planning, analysis and evaluation, while the main thrust of SE is on core design aspects.

**Fig 1 IS and SE Foci on Systems Development Life-Cycle**

Although the emphasis in the disciplines might differ, it is certainly the case that the topic coverage between the disciplines is closer than say a comparison of IS with a mainstream management journal, or SE with a mainstream computer science journal.

There seems to be a growing recognition in the Software Engineering field that an enlargement of scope to consider human factors is necessary. This is evident, for example, in the recent special section of *TSE* on socio-technical environment of software development projects (*TSE* May/June 2011). Socio-technical issues have long been a feature of IS research. Likewise research method issues are now looming larger in SE. This is evident in guidelines on systematic literature reviews, for example, and is also a theme that fits well with the empirical software engineering domain. Again, a (perhaps excessive) focus on research methods has long been a part of IS research.

**Fusing IS and SE Insights in Design[3]**
Design is a prominent research topic in both disciplines. The roots of the IS discipline are that of science of the artificial which is deeply engaged in the design of the artificial world. A key challenge in design is to design for mutability, that is systems which can evolve to cope with changing contexts without a loss of functionality or quality. Both the IS and SE fields contribute important insights to the design challenge. Taking the case of IS, for example, there is a widely-held acceptance that it is impossible to completely specify how a system will be used in advance, as emergent properties between the technological and social worlds ensure that intended use and 'real' use are not one and the same (Keen & Scott-Morton, 1978; Orlikowski 1992). The fact that technological artifacts can transform the social world in which they are implemented is abundantly clear in the face of social media such as Facebook, Twitter, LinkedIn, or game/virtual world contexts such as Second Life which have now been legitimated as business tools (*Business Week* 2006).

A useful typology of mutability in design is proposed by Sjostrom et al (2011) who propose a distinction between mutability-in-design and mutability-in-use. The differentiation between design and use has already been mentioned, and has long been recognized in IS and HCI disciplines.

Taking the mutability-in-design perspective initially, this is an area where SE provides much by way of useful insights. This is evident in early research by Parnas on information-hiding and 'designing for change' (Parnas 1976). Separation of concerns leads to removal of dependencies. Software architecture and software patterns represent initiatives which seek to separate business logic from user interface logic – the model-view-controller (MVC) pattern for example. Furthermore, automated testing, becoming increasingly possible, helps reduce the risk of errors caused by software adaptation, and again a number of basic software patterns help facilitate automated testing – the Law of Demeter and Inversion of Control, for example.

Moving to the mutability-in-use perspective, IS contributes more significantly. This requires an understanding of the different system users and recognizes the need for a configurable system which can adapt to different stake-holder needs, while also accepting that there is a design trade-off between simplicity and configurability (Sjostrom et al 2011). From an SE perspective, Jackson (1995) usefully considers the world and machine perspectives and the importance of design in context. He suggests the primary concern for a system is that of providing practical value in the real world, not one of code structure *per se*. However the latter is clearly important to facilitate change so that a designer can identify what might change, that system can be aware of it, and continue to work after such changes have been made. Furthermore, IS recognizes the unintended ways in which technology may be adapted in use. The manner in which SMS texting was transformed from a technology conceived as being aimed at high cost commercial exchanges transacted by corporate professionals to its eventual use as a low cost mass market medium for informal communication is evidence of this.

---

[3] This section draws largely on Sjostrom et al 2011

## Towards Practical Collaboration

The first step in addressing any challenging situation is to accept the challenges and recognize potential opportunities to make progress. As stated at the outset, the purpose of this article is to encourage the process of interaction between the IS and SE fields.

Interestingly, the perception of the other field does not appear to be symmetric in both cases. IS research frequently cites SE research and uses research concepts quite freely from SE. The reverse does not appear to be the case as SE research more rarely cites IS research. Evidence of the relatively closed and inward focus of SE was presented in Glass et al (2004) who reported that only 1.9% of the SE papers used theories and models from other disciplines. For comparison, computer science papers used other disciplines in 10.77% of the cases, whilst Information Systems papers used other disciplines in 67.9% of the cases. This is in keeping with Davis' intersection approach to establishing conceptual foundations.

Even the SWEBOK – which has been criticized for being too inclusive – lists seven disciplines as related to SE – cognitive science and human factors; computer engineering; computer science; management and management science; mathematics; project management and systems engineering, It does not include IS as a related discipline, even though the topic overlap in the journals presented above is clear evidence of a relationship. In a similar fashion, however, Davis' (2000) identification of "underlying disciplines" for IS mentions psychology, sociology, economics, and systems concepts and principles, but, tellingly, not software engineering.

Anecdotally, the lack of receptivity by SE to IS research is illustrated by a case where the main reason for rejection of a paper by a prominent SE journal was that it was "an IS paper".

There are some exceptions. For example, Vic Basili, a prominent figure in the SE field, was an invited speaker at the ICIS 2007 conference where he spoke about bridging the gap between business strategy and software development, a topic of central importance in the IS field (Fitzgerald 1998; Fitzgerald et al 2000). However such cross-over participations are rare. For the benefit of the ICT field at large, and the society it exists to serve, we need to break down such barriers and achieve mutual respect and fruitful collaboration.

A number of practical initiatives could be easily established to facilitate potential convergence across both fields. To help stimulate discussion on the issue, panel sessions at the respective conferences in each discipline could discuss some of the issues involved. Conferences in each field could establish a track to showcase relevant research from the other discipline. This could be extended to special sections in the journals in each discipline.

Replication studies are the heart of science. A useful initiative could be for IS and SE researchers to try replicate studies from the other discipline to generate fresh insights into open questions.

Scientific fields are no longer coterminous with academic disciplines (Whitley 2000). Thus, the danger for each field is that their traditional research ground could be usurped by other disciplines. Combining insights from both fields would lead to a more solid basis for each. It has been observed that a scientist would rather use someone else's toothbrush than another scientist's nomenclature. It is surely time to swap toothbrushes!

**Acknowledgements**

I would like to acknowledge the useful input from Par Agerfalk, Bashar Nuseibeh and Kevin Ryan who provided feedback and stimulated my thinking on this topic.

**References**

Russell L. Ackoff, "Management Misinformation Systems," *Management Sciences* 14, no. 4 (December 1967). The Institute of Management Sciences, 290 Westminster Street, Providence, R.I. 02903.

Davis, G. B. *Management Information Systems: Conceptual Foundations, Structure and Development*. New York: McGraw-Hill Book Company, 1974.

Davis, G.B. (2000) Information systems conceptual foundations: looking backward and forward, in Baskerville, R, Stage J and DeGross J (Eds) *Organizational and Social Perspectives on Information Technology, Proceedings of the IFIP TC8 WG8.2 international conference*, Aalborg, Denmark.

Dijkstra, E (1968) GoTo Statement Considered Harmful, *Communications of the ACM*, Vol. 11, No. 3, pp. 147-148

Fitzgerald, B. (1998). An empirically-grounded framework for the information systems development process.

Fitzgerald, B., Russo, N., & O'Kane, T. (2000). An empirical study of system development method tailoring in practice. *ECIS 2000 Proceedings*, 4.

Glass, R, Ramesh, B and Vessey, I (2004) *Communications of ACM*
IEEE Standard Glossary of Software Engineering Terminology, IEEE std 610.12-1990, 1990.

Keen, P. and Scott Morton, M. (1978) *Decision Support Systems: An Organizational Perspective*, Addison-Wesley, Reading.

Langefors B. (1966). *Theoretical Analysis of Information Systems.*Lund: Studentlitteratur.

Naur, P., and Randell, B. (eds.) (1969) *Software Engineering: A Report on a Conference Sponsored by the NATO Science Committee*. Brussels: Scientific Affairs Division, NATO.

Orlikowski, W (1992) The Duality of Technology: Rethinking the Concept of Technology in Organizations, *Organization Science*, 3, 3, 398-427.

Parnas, D. (1972) On the criteria to be used in decomposing systems into modules. *Communications of the ACM*, **15**, 12, 1053-1058.

Parnas, D. (1976) On the design and development of program families. *IEEE Transactions on Software Engineering,* Vol. SE-2, pp.1-9

Sidorova, A., Evangelopoulos, N., Valacich, J.S., and Ramakrishnan, T. "Uncovering the intellectual core of the information systems discipline," *MIS Quarterly* (32:3) 2008, pp 467–482

Sjostrom, J, Agerfalk, P and Lochan, R (2011) Mutability matters: baselining the consequences o0f design, *MCIS Conference,* Cyprus.

Ward, P. (1991) The evolution of structured analysis: Part I--the early years. *American Programmer*, **4**, 11, 4-16.

Whitley, R. (2000) *The Intellectual and Social Organization of the Sciences,* Oxford: Oxford University Press.